\documentclass[aps,pra,reprint,nofootinbib,superscriptaddress]{revtex4-1}

\usepackage{amsmath}
\usepackage{amsfonts}
\usepackage[amssymb]{SIunits}
\usepackage{url}
\usepackage{graphicx}
\usepackage{xcolor}

\usepackage{soul}

\usepackage{hyperref}
\usepackage{mathtools}
\usepackage{rotating}
\usepackage{units}
\usepackage{lipsum}
\usepackage{relsize} 
\usepackage{nicefrac}

\begin{document}

\title{Spectrally resolved linewidth enhancement factor of a semiconductor\\ frequency comb}

\author{Nikola~\surname{Opa\v{c}ak}}
\email{nikola.opacak@tuwien.ac.at}
\affiliation{Institute of Solid State Electronics, TU Wien, Vienna, Austria}
\author{Florian~\surname{Pilat}}
\affiliation{Institute of Solid State Electronics, TU Wien, Vienna, Austria}
\author{Dmitry~\surname{Kazakov}}
\affiliation{Institute of Solid State Electronics, TU Wien, Vienna, Austria}
\affiliation{John A. Paulson School of Engineering and Applied Sciences, Harvard University, Cambridge, Massachusetts USA}
\author{Sandro~\surname{Dal~Cin}}
\affiliation{Institute of Solid State Electronics, TU Wien, Vienna, Austria}

\author{Georg~\surname{Ramer}}
\affiliation{Institute of Chemical Technologies and Analytics, TU Wien, Vienna, Austria}

\author{Bernhard~\surname{Lendl}}
\affiliation{Institute of Chemical Technologies and Analytics, TU Wien, Vienna, Austria}

\author{Federico~\surname{Capasso}}
\affiliation{John A. Paulson School of Engineering and Applied Sciences, Harvard University, Cambridge, Massachusetts USA}

\author{Benedikt~\surname{Schwarz}}
\email{benedikt.schwarz@tuwien.ac.at}
\affiliation{Institute of Solid State Electronics, TU Wien, Vienna, Austria}
\affiliation{John A. Paulson School of Engineering and Applied Sciences, Harvard University, Cambridge, Massachusetts USA}

\begin{abstract}
The linewidth enhancement factor (LEF) has recently moved into the spotlight of research on frequency comb generation in semiconductor lasers.
Here we present a novel modulation experiment, which enables the direct measurement of the spectrally resolved LEF in a laser frequency comb.
By utilizing a phase-sensitive technique, we are able to extract the LEF for each comb mode.
We first investigate and verify this universally applicable technique using Maxwell-Bloch simulations and then present the experimental demonstration on a quantum cascade laser frequency comb.
\end{abstract}

\maketitle

Semiconductor lasers are compact, electrically pumped and provide substantial broadband gain. They are recently gaining vast attention due to a wide range of applications that utilize their coherence properties, such as high-precision spectroscopy~\cite{haensch2006nobel}.
Their asymmetric gain spectrum additionally sets them apart from other laser types, where the lasing transition takes place between two discrete levels. 
Following the Kramers-Kronig relations, an asymmetric gainshape results in a dispersion curve of the refractive index that has a non-zero value at the gain peak, where the emission frequency of a free-running laser lies~\cite{yariv1989quantum}. As a consequence, a remarkable property of semiconductor lasers is that both the refractive index and the optical gain change simultaneously with the varying carrier population~\cite{Osinski1987Linewidth}. 
This was quantified with the linewidth enhancement factor (LEF), also called the $\alpha$-factor, defined by Henry as the ratio of changes of the modal index and gain at the gain peak~\cite{Henry1982Theory}.
Many unique properties of semiconductor lasers can be traced back to the non-zero value of this factor.
The LEF was first introduced in the 1980s to describe the broadening of the semiconductor laser linewidth~\cite{Henry1982Theory,vahala1983semiclassical} beyond the Schawlow-Townes limit~\cite{schawlow1958infrared}.
Furthermore, the LEF determines the dynamics of semiconductor lasers as it describes the coupling between the amplitude and phase of the optical field~\cite{agrawal1995semiconductor,Gray1992Importance}.
In lasers with fast gain recovery times, the LEF was recently connected to the onset of a giant Kerr nonlinearity~\cite{opacak2021frequency} and frequency modulated combs~\cite{opacak2019theory}.
It was shown that the LEF can lead to a low-threshold multimode instability and frequency comb formation~\cite{piccardo2020freqeuncy,meng2020midIR}. Appropriate values of the LEF were predicted to result in the emission of solitons in active media~\cite{columbo2020unifying}. 
 The precise knowledge of the LEF represents a key point in understanding many astonishing features of semiconductor lasers and subsequently controlling them.

The physical origin of the LEF is explained through the asymmetric gain spectrum of semiconductor lasers. In interband lasers, this is due to the opposite curvatures of the valence and conduction bands in k-space~\cite{agrawal1995semiconductor}, which yield LEF values around 2-7~\cite{Osinski1987Linewidth}. In intersubband lasers, the states have similar curvatures, so the gain asymmetry originates from the non-parabolicity~\cite{liu2013importance}, counter-rotating terms~\cite{pereira2016linewidth}, and Bloch gain~\cite{opacak2021frequency}. Measured values range from -0.5 to 2.5~\cite{Aellen2006direct,jumpertz2016measurements,hangauer2015gain,vonStaden2006measurement,piccardo2020freqeuncy,piccardo2020freqeuncy}.

An established technique for extracting subthreshold values of the LEF is the Hakki-Paoli method by measuring the gain and wavelength shift~\cite{hakki1975gain,henning1983measurement}. Above threshold values can be inferred from the measurements of the linewidth broadening~\cite{Henry1982Theory} and the phase noise~\cite{henry1983theory}.
Other methods are based on the analysis of the locking regimes induced by optical injection from a master laser~\cite{liu2001measurement}, or on the optical feedback and the characterization of the self-mixing signal~\cite{yu2004measurement,fan2015simple}.
Harder et al. provided a study of the laser's response under a modulation of the injection current and were able to extract the LEF value~\cite{harder1983measurement}. A modified experiment which includes heterodyning allowed a direct measurement of the LEF~\cite{Aellen2006direct}.
However, all of the mentioned techniques have one substantial limitation -- they do not resolve the spectral behavior of the LEF.
Most methods rely on single-mode operation, which is either achieved in Fabry-P\'{e}rot lasers very close to threshold or using distributed feedback (DFB) laser. In DFBs, the lasing wavelength is detuned from the exact gain peak, thus introducing an uncertainty in the LEF measurement.

\begin{figure*}[htb!]
	\centering
	\includegraphics[width = 1 \textwidth]{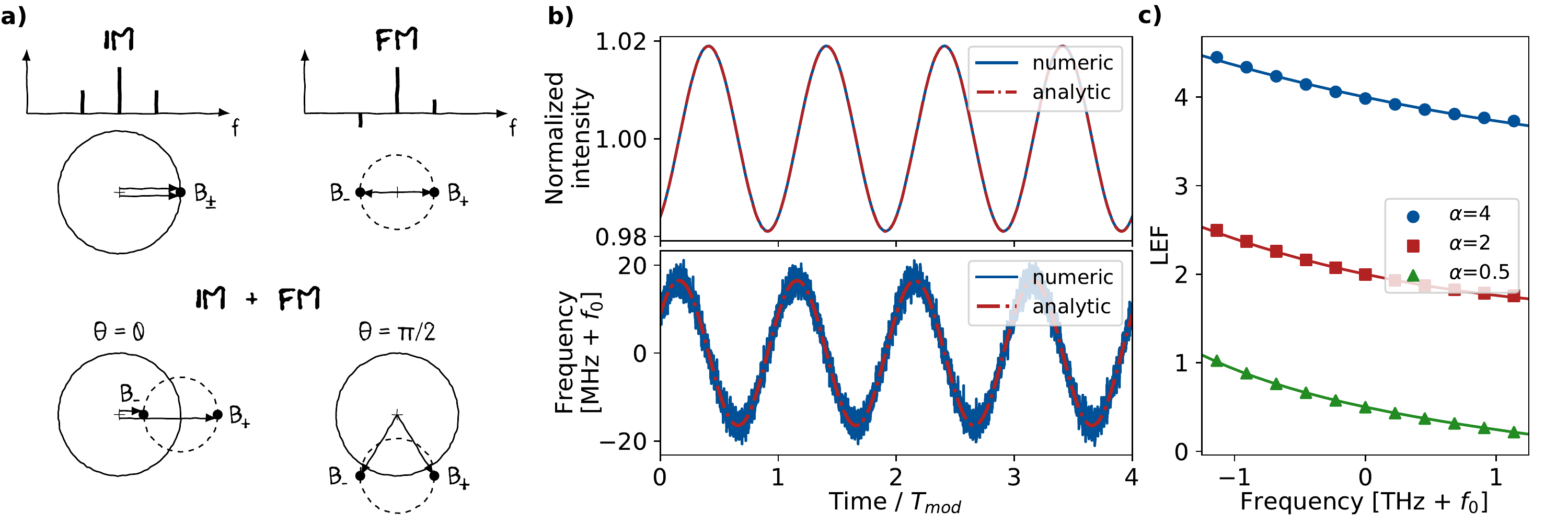}
	\caption{ \textbf{Intensity modulation (IM) and frequency modulation (FM) of a single-mode laser and the LEF extraction. } \textbf{(a)} Modulation sidebands and their beating signals $B_{\pm }$ with the center mode in the case of a pure IM or FM (top). The beatings are represented in the complex plane. A mixture of both IM and FM is analysed below for two values of the IM-FM phase shift $\theta$. \textbf{(b)} Time traces of the intensity and frequency. Analytic curves given by Eq.~(\ref{eqm:1}) (red dash-dotted lines) are fitted to the numeric time traces (blue solid lines), obtained from a simulation~\cite{opacak2019theory}. The frequency is modulated around the lasing frequency $f_0$ and the time is normalized to the modulation period $T_\mathrm{mod}$, with $f_\mathrm{mod}=3~\mathrm{GHz}$. The phase shift between the frequency and intensity modulations is $\theta \approx \pi/2$.  \textbf{(c)} Calculated LEF of a simulated single-mode laser. The lasing frequency is swept in steps around the gain peak frequency $f_0$. The extracted values match the LEF from an analytic model (solid lines).  }
	\label{fig1}
\end{figure*}

In this work, we introduce a novel measurement technique, which enables the direct and spectrally resolved measurement of the LEF of a semiconductor laser operating in the frequency comb regime.
It builds upon the modulation experiment by Harder et al.~\cite{harder1983measurement} and enables the extraction of the LEF for each individual comb mode from a single measurment.
This is made possible by the shifted-wave interference Fourier transform spectroscopy (SWIFTS)~\cite{burghoff2015evaluating}. Each comb mode, together with its neigboring modulation sidebands, produces beatings. SWIFTS allows the spectrally resolved measurement of both the amplitude and phase of these beatings. 
We first lay down the theoretical foundations of our method. The analytic values of the LEF are subsequently compared with the values extracted from a numerical model of both a single-mode and multimode laser. This is followed by the experimental demonstration, where the method is employed on a quantum cascade laser frequency comb.

In the presence of a non-zero value of the LEF, a sinusoidal modulation of the bias current results in both an intensity modulation (IM) and a frequency modulation (FM) of the laser output~\cite{agrawal1995semiconductor,harder1983measurement}. Similarly to the work done in~\cite{zhu1997modulation}, the light emitted by a single-mode laser with an average intensity $I_0$ is described with an electric field of:

\begin{align}
\begin{split}
E(t) = &\sqrt{I_0}\sqrt{1 +m \cos{(2\pi f_\mathrm{mod} t + \phi)}} \\
&\times \cos{ \big( 2\pi f_0 t + \beta \sin{(2\pi f_\mathrm{mod} t + \phi + \theta)}    \big) },
\label{eqm:1}
\end{split}
\end{align}

where $f_0$ is the lasing frequency, $f_\mathrm{mod}$ is the modulation frequency, $m$ and $\beta$ are the IM and FM indices, respectively. We include an additional arbitrary phase shift $\phi$ with respect to the current modulation and an FM-IM phase shift $\theta$. Following Eq.~(\ref{eqm:1}), the frequency is modulated as $f=f_0+\beta f_\mathrm{mod} \cos(2\pi f_\mathrm{mod} t + \phi + \theta)$.

\begin{figure}[b!]
	\centering
	\includegraphics[width = 1\columnwidth]{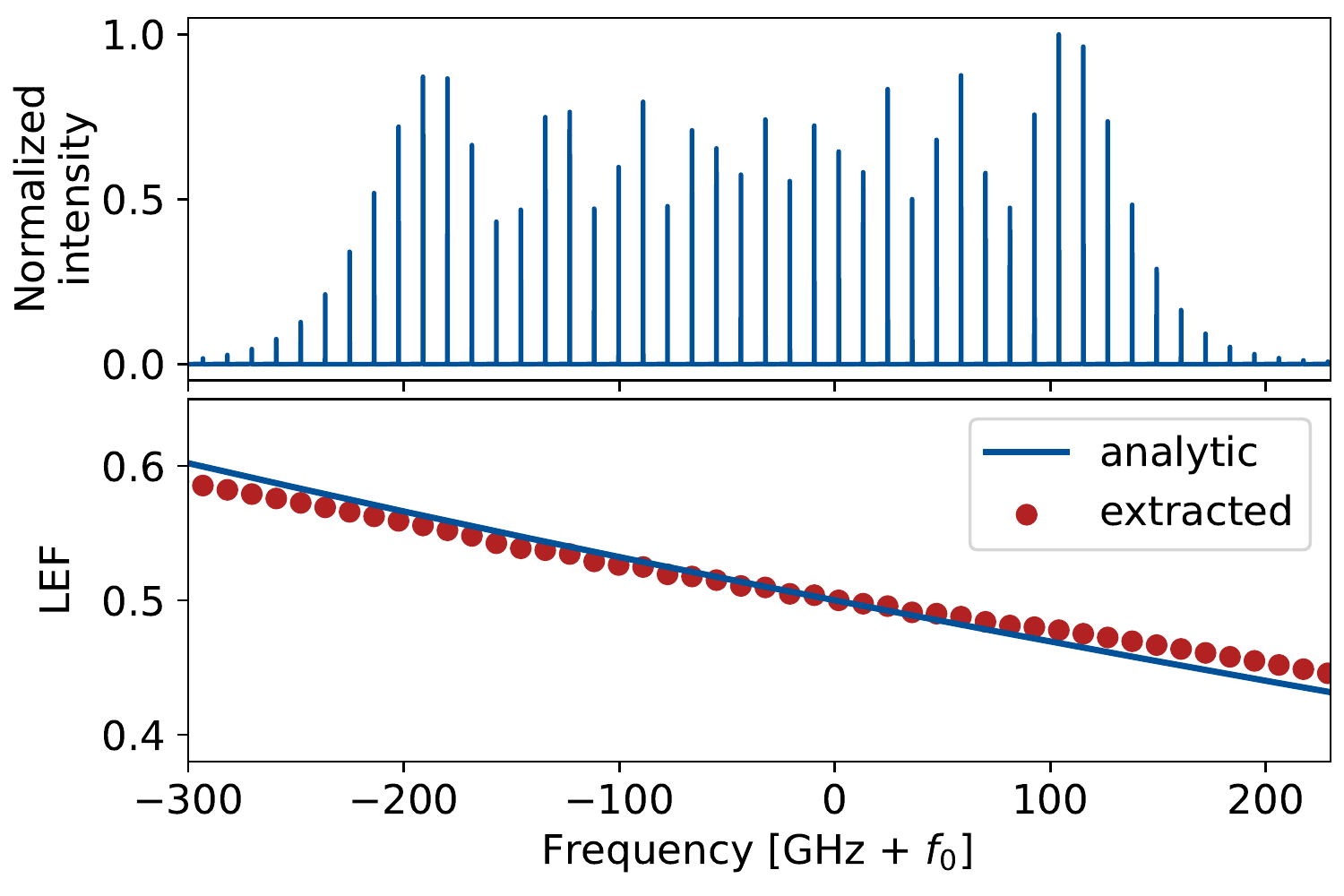}
	\caption{ \textbf{LEF of a simulated multimode semiconductor laser. } The normalized intenisty spectrum is depicted on the top. The spectrally resolved LEF is shown on the bottom. Red dots represent the LEF for each lasing mode, calculated using Eq.~(\ref{eqm:4}). The solid blue line represents the LEF from the analytic model, given by the Eq.~(B3) in the Appendix.  }
	\label{fig2}
\end{figure}

\begin{figure*}[th!]
	\centering
	\includegraphics[width = 1 \textwidth]{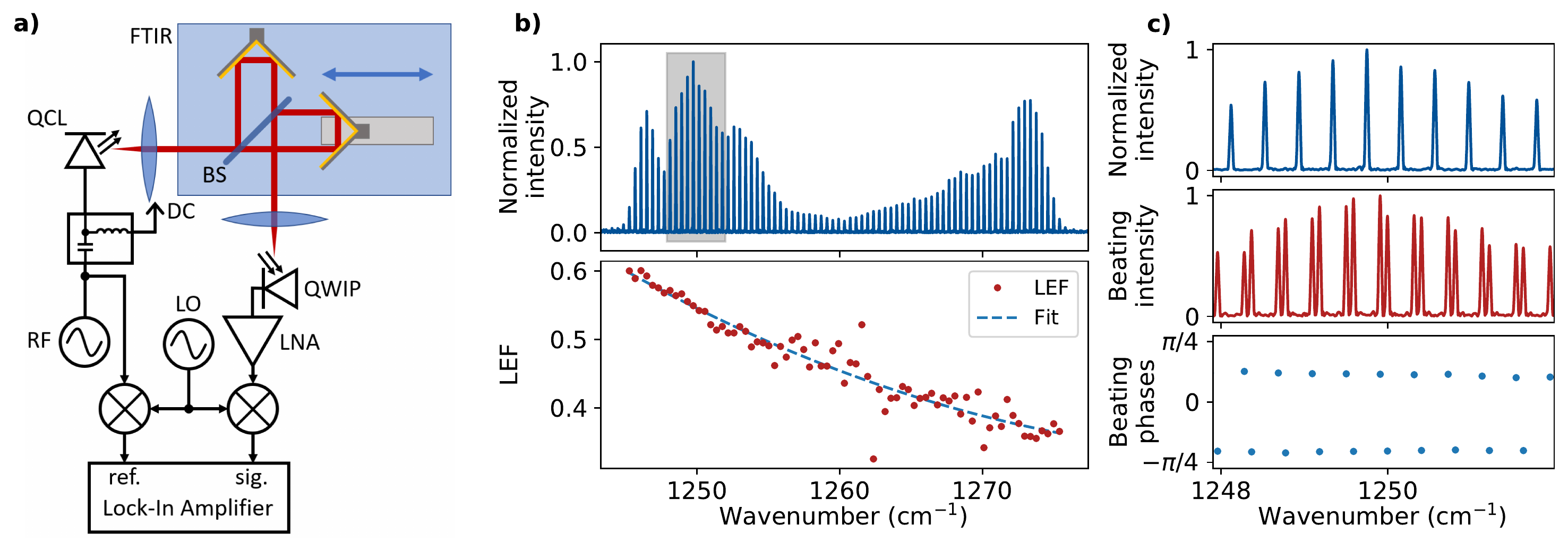}
	\caption{ \textbf{Experimental data of a SWIFTS measurement with LEF evaluation. } \textbf{(a)} Sketch of the experimental setup: Fourier transform infrared spectrometer (FTIR), beam splitter (BS), quantum cascade laser (QCL), bias voltage (DC), RF generator (RF), local oscillator (LO), quantum well infrared photodetector (QWIP), low-noise amplifier (LNA). \textbf{(b)} Measured intensity spectrum of the QCL and extracted LEF values for each mode (red dots) with a fit (blue dashed line). \textbf{(c)} Intensity (top), beating intensity (middle) and beating phase spectrum as obtained from the SWIFTS measurement for the highlighted grey area in (b).}
	\label{fig3}
\end{figure*}

Using the Jacobi-Anger expansion~\cite{abramowitz1965handbook}, Eq.~(\ref{eqm:1}) can be written as a Fourier series:

\begin{align}
E(t)= \sqrt{I_0} \sum_{n=-\infty}^{+\infty} E_n   \exp{ \big( 2\pi (f_0 + n f_\mathrm{mod}) t \big)  }  ,
\label{eqm:2}
\end{align}
as is derived in the Supplementary Material. 
Under the assumption of a weak modulation strength ($ m,\beta \ll1 $), the complex beating signals $B_{\pm }$ between the centerline $E_0$ and its first modulation sidebands $E_{\pm 1}$ can be written as:

\begin{align}
\begin{split}
B_{+} &= E_{1}E^*_0 = \mathrm{e}^{  i (\phi + \theta ) } \Big( \frac{\beta}{2} + \frac{m}{4} \mathrm{e}^{ - i \theta } \Big) \\
B_{-} &= E_{0}E^*_{-1} = \mathrm{e}^{  i (\phi + \theta ) } \Big(- \frac{\beta}{2} + \frac{m}{4} \mathrm{e}^{ - i \theta } \Big).
\label{eqm:3}
\end{split}
\end{align}

The extraction of the modulation indices $m$ and $\beta$ is possible from Eq.~(\ref{eqm:3}), if both the amplitudes and phases of $B_\pm$ are known.
Furthemore, this is also valid in the case of a multimode laser, where each mode $k$ produces beatings $B_{k,\pm}$ with its neigboring modulation sidebands.
With the knowledge of the modulation indices, the spectral LEF for each mode can finally be calculated directly as~\cite{harder1983measurement,Osinski1987Linewidth,Aellen2006direct,hangauer2015gain}: 

\begin{align}
\mathrm{LEF}_k= 2\frac{\beta_k}{m_k} = \bigg|  \frac{B_{k,+}-B_{k,-}}{B_{k,+}+B_{k,-}}  \bigg| ,
\label{eqm:4}
\end{align}

The modulation sidebands of a single-mode laser are sketched in Fig.~\ref{fig1}a (top) in the cases when only IM or FM is present. The corresponding beating signals $B_{\pm}$ are plotted below in the complex plane, following the analysis of Eq.~(\ref{eqm:3}). They are in-phase with each other in the case of a pure IM, and anti-phase ($\pi$~phase-shifted) for a perfect FM. In a modulated semiconductor laser, a mixture of IM and FM is always present. The beating signals in this case are sketched on the bottom of Fig.~\ref{fig1}a for two exemplary values of the FM-IM phase shift $\theta=0$ and $\pi/2$, which is unknown a priori.
This highlights why a phase-sensitive technique is required to measure the amplitudes and phases of $B_\pm$.

In Fig.~\ref{fig1}b, we show the instantaneous intensity and frequency of a semiconductor single-mode laser biased above threshold with a small superimposed sinusoidal modulation. The results are obtained from a numerical time-domain model of the laser based on the Maxwell-Bloch formalism~\cite{opacak2019theory}. The analytical model given by Eq.~(\ref{eqm:1}) is fitted to the numerical time traces of both the instantaneous intensity and frequency. The IM-FM phase shift approaches ${\pi}/{2}$~\cite{hangauer2014high,hangauer2015gain}, as can be seen from Fig.~\ref{fig1}b. Fig.~\ref{fig1}c shows the calculated LEF of a simulated single-mode laser, whose frequency was tuned in discrete steps around the constant gain peak frequency $f_0$. We show the results for three different values of the LEF at the gain peak $\alpha=0.5$, $2$ and $4$, represented by green triangles, red squares and blue circles, respectively. The extracted LEF values from the numerical model follow the LEF obtained from the analytic model, which is plotted with solid lines. The analytic model is given by Eq.~(B3) in the Appendix.

In the following we will discuss how to employ this technique to a frequency comb. 
The modulation of the current induces modulation sidebands around each comb mode. By finding the beatings $B_{k,\pm}$ of each mode with its corresponding neighboring sidebands, one can obtain the LEF for each comb mode. Fig.~\ref{fig2} shows an intensity spectrum of a simulated laser in a multimode comb regime. The extracted spectrally resolved LEF is plotted with red dots below, together with the LEF from an analytic model of the laser gain medium (Eq.~(B3) in the Appendix), represented with the blue solid line. 
The slight deviations can be attributed to coherent mechanisms that couple the frequency comb modes.

In experiments, the amplitudes and phases of the beatings $B_{k,\pm}$ can be measured most elegantly using SWIFTS.
The experimental setup is shown in Fig.~\ref{fig3}a.
SWIFTS employs a Fourier-transform infrared (FTIR) spectrometer in order to spectrally resolve all individual beatings.
The modulation of the laser is measured in amplitude and phase using a fast photodetector and a lock-in amplifier to obtain the SWIFTS interferogram (see Fig.~C1 in the Appendix).
The beatings are obtained using the Fourier transformation.
We extended our previous SWIFTS setup~\cite{hillbrand2018coherent} with a custom-built high-resolution FTIR spectrometer ($\approx$500$\,$MHz), in order to resolve the narrowly spaced beatings. The setup consists of a Newport Optical Delay Line Kit (using DL325), a broadband mid-infrared beamsplitter, a temperature stabilized HeNe-Laser and a Zürich Instruments HF2LI lock-in amplifier for the acquisition of the intensity and SWIFTS interferograms of the measured laser, as well as the interferogram of the HeNe-laser. We operate the FTIR spectrometer in rapid scan mode and employ the Hilbert-transform to obtain the mirror delay from the HeNe interferogram.

The QCL, which was used in this measurement, is a $ 3.5\,$mm long ridge laser emitting at around $8\,\mathrm{\mu}$m.
It consists of a long gain section and a shorter modulation section to efficiently couple RF signals to the laser. This feature, however, is not a necessity for the proposed technique. The laser was operated in a free-running frequency comb state with a repetition frequency of 12.2 GHz. The frequency of the weak modulation was chosen to be sufficiently lower at 9.593 GHz and a demodulation frequency of 9.570 GHz was set on the local oscillator.
An RF-optimized quantum well infrared photodetector (QWIP), cooled to 78 K, was used for light detection.  
The intensity spectrum (Fig.~\ref{fig3}b) shows a frequency comb spanning over 75 modes. A zoom of the grey-shaded area can be seen in Fig.~\ref{fig3}c. The weak modulation sidebands are below the noise of the intensity spectrum (top).
Nevertheless, the beatings, produced by the laser modes and their weak sidebands, can be precisely measured using SWIFTS in both the amplitudes and the phases (middle and bottom, respectively).  
Applying Eq.~\ref{eqm:4} to the beating amplitudes and phases, we extract the LEF for each individual comb mode, as displayed in Fig. \ref{fig3}b (bottom). The extracted spectrally resolved LEF follows the expected characteristic shape (see Fig.~\ref{fig1}c and \ref{fig2}). Its value matches the prediction from recent theoretical work, which pinpointed the origin of the LEF to the Bloch gain~\cite{opacak2021frequency}.

In conclusion, we present a novel technique to directly measure the spectrally resolved LEF of a running semiconductor laser frequency comb.
%Our method will allow to extract the spectral LEF in frequency combs based on any type of a semiconductor laser. 
The measurement concept is first verified using elaborate numerical simulations of a modulated semiconductor laser. There, an excellent agreement is observed with the expected spectral LEF from the analytic model. 
The experimental demonstration was performed on a multimode QCL frequency comb, while the technique itself is universal.
It will allow to extract the spectral LEF in frequency combs based on any type of a semiconductor laser. 
The LEF governs many coherent processes in a running semiconductor lasers among which is the frequency comb regime. Its precise knowledge will provide a better fundamental understanding of the light evolution that will promote further technological advancements.

This work was supported by the Austrian Science Fund (FWF) within the projects ”NanoPlas” (P28914-N27) and by the European Research Council (ERC) under the European Union’s Horizon 2020 research and innovation programme (Grant agreement No. 853014).

\footnotesize

\providecommand{\noopsort}[1]{}\providecommand{\singleletter}[1]{#1}

\bibliography{mybib}

\normalsize

%%%%%%%%%%%%%%%%%%%%%%%%%%%%%%%%%%%%%%%%%%%%%%%%%%%%%%%%%%%%%%%%%%%%%%%%%%%%%%%%%%%%%%%%%%%%%%%%%%%%%%%%%
\appendix
\renewcommand{\thefigure}{C\arabic{figure}}
\setcounter{figure}{0}
%\pagebreak
%{\huge Supplementary material}
%\section{\huge Supplementary material}

\section{Theoretical background}
%\subsection{Theoretical background}

It is known from literature that the width of the laser line is due to the fluctuations of the phase of the optical field~\cite{Henry1982Theory}. These fluctuations partly come from the spontaneous emission events within the laser active medium, which alter the phase and the intensity of the laser field. This, together with the resonator losses, defines the Schawlow-Townes limit of the laser linewidth~\cite{schawlow1958infrared}. However, semiconductor lasers are known to possess a much larger linewidth~\cite{Osinski1987Linewidth}. This was explained as a consequence of an asymmetric spectral shape of the available optical gain. Following the Kramers-Kronig relations, an asymmetric gain shape will induce a non-zero value of the refractive index dispersion line at the position of the gain peak. Since the gain depends on the carrier density, the refractive index will change as the carrier density changes. The change of the refractive index in a limited amount of time will enforce an additional phase shift and an additional broadening of the linewidth which is much greater than the one induced by the spontaneous emission. The dynamic coupling of the refractive index and gain is described simply by the ratio of their respective carrier-induced changes. This ratio is known as the linewidth enhancement factor (LEF), or the $\alpha$-factor. In this work we will reveal for the first time a measurement technique that is able to extract the value of LEF over the whole spectrum of an operating laser.

We will start by analyzing the response of a single-mode semiconductor laser biased above threshold with a superimposed weak modulation. The bias current $J(t)$ is modulated sinusoidally as:

\begin{flalign}
\begin{split}
	J(t) = J_0 + { \Delta} J \cos{(2\pi f_\mathrm{mod} t)},
\label{eq:1} 
\end{split}
\end{flalign}

where $f_\mathrm{mod}$ is the modulation frequency. Due to the non-zero LEF in semiconductor lasers, a modulation of the current will result in the modulation of both the laser intensity and frequency. The optical field that is emitted by the laser is the defined as:

\begin{align}
\begin{split}
E(t) = & \sqrt{I_0}\sqrt{1 +m \cos{(2\pi f_\mathrm{mod} t + \phi)}} \\
 &\times \cos{ \big( 2\pi f_0 t + \beta \sin{(2\pi f_\mathrm{mod} t + \phi + \theta)}    \big) },
\label{eq:2}
\end{split}
\end{align}

where $I_0$ is the average intensity, $f_0$ is the lasing frequency, and $m$ and $\beta$ are the intensity modulation (IM) and frequency modulation (FM) indices, respectively. We include an additional arbitrary phase shift $\phi$ with the respect to the current modulation and an FM-IM phase shift $\theta$. Following Eq.~(\ref{eq:2}), the frequency is modulated as $f=f_0+\beta f_\mathrm{mod} \cos(2\pi f_\mathrm{mod} t + \phi + \theta)$. For the following calculation it is of great use the complex field notation and assume weak modulation ($m,\beta \ll 1$):

\begin{align}
\begin{split}
E(t) = &\sqrt{I_0}{\big(1 + \frac{m}{2} \cos{(2\pi f_\mathrm{mod} t + \phi)}\big)} \\
 &\times \mathrm{e}^{ i 2\pi f_0 t + i \beta \sin{(2\pi f_\mathrm{mod} t + \phi + \theta)}     }.
\label{eq:3}
\end{split}
\end{align}

Since the modulation of $E(t)$ is a periodic function with frequency $f_{\mathrm{mod}}$, we can expand Eq.~(\ref{eq:3}) in an infinite Fourier series of harmonic terms that oscillate at frequencies $f_0 + n f_{\mathrm{mod}}, n \in \mathbb{Z}$. It follows:

\begin{align}
E(t)= \sqrt{I_0} \sum_{n=-\infty}^{+\infty} E_n   \exp{ \big( 2\pi (f_0 + n f_\mathrm{mod}) t \big)  }  ,
\label{eq:4}
\end{align}

where the complex amplitudes $E_n$ are to be determined. This is feasible by employing the Jacobi-Anger expansion~\cite{abramowitz1965handbook}:

\begin{align}
\mathrm{e}^{i x \sin{y}} = \sum_{n=-\infty}^{+\infty} J_n(x) ~\mathrm{e}^{iny},
\label{eq:5}
\end{align}

where $J_n(x)$ is the $n$-th Bessel function of the first kind. Combining Eqs.~(\ref{eq:3}) and (\ref{eq:5}), using the identity $J_{-n}(x)=(-1)^n J_n(x)$ for $n \in \mathbb{Z}$, and keeping only the orders $n= -2,-1,0,1,2$ yields:

\begin{align}
\begin{split}
E(t) &= \sqrt{I_0} ~\mathrm{e}^{i 2 \pi f_0 t}\Big( 1 +\frac{m}{4} \mathrm{e}^{i(2\pi f_\mathrm{mod} t + \phi)} \\ & +\frac{m}{4} \mathrm{e}^{-i(2\pi f_\mathrm{mod} t + \phi)} \Big) 
 \Big[ 
 J_2(\beta)~\mathrm{e}^{ -i2 {(2\pi f_\mathrm{mod} t + \phi + \theta)}}  \\
 &-J_1(\beta)~\mathrm{e}^{ -i {(2\pi f_\mathrm{mod} t + \phi + \theta)}} 
  +J_0(\beta) \\
 &+ J_1(\beta)~\mathrm{e}^{ i {(2\pi f_\mathrm{mod} t + \phi + \theta)}}    +  J_2(\beta)~\mathrm{e}^{ i2 {(2\pi f_\mathrm{mod} t + \phi + \theta)}}\Big].
\label{eq:6}
\end{split}
\end{align}

Comparison of Eqs.~(\ref{eq:4}) and~(\ref{eq:6}) yields the expressions for the complex amplitude of the centerline $E_0$ and its first modulation sidebands $E_{\pm1}$: 

\begin{align}
\begin{split}
E_0 & = 1 \\
E_{\pm1} &= \mathrm{e}^{ \pm i (\phi + \theta ) } \Big( \pm J_1(\beta) + \frac{m}{4} \big(  J_2(\beta )\mathrm{e}^{ \pm i \theta } + J_0(\beta )\mathrm{e}^{ \mp i \theta } \big) \Big).
\label{eq:7}
\end{split}
\end{align}

We can now also employ the weak modulation approximation so that $J_0(\beta) \approx 1$, $J_1(\beta) \approx \frac{\beta}{2}$, and $J_2(\beta) \approx 0$ holds. By finding the beatings between the centerline and its first sidemodes, we can directly extract the modulation indices $m$ and $\beta$. The beating signals $B_{\pm 1}$ are defined as:

\begin{align}
\begin{split}
B_{1} &= E_{1}E^*_0 = \mathrm{e}^{  i (\phi + \theta ) } \big( \frac{\beta}{2} + \frac{m}{4} \mathrm{e}^{ - i \theta } \big) \\
B_{-1} &= E_{0}E^*_{-1} = \mathrm{e}^{  i (\phi + \theta ) } \big(- \frac{\beta}{2} + \frac{m}{4} \mathrm{e}^{ - i \theta } \big).
\label{eq:8}
\end{split}
\end{align}

The LEF can be calculated from the modulation indices, as is known from literature~\cite{harder1983measurement,Osinski1987Linewidth,Aellen2006direct,hangauer2015gain}:

\begin{align}
\mathrm{LEF} = 2 \frac{\beta}{m}  .
\label{eq:9}
\end{align}

Combining Eqs.~(\ref{eq:8}) and~(\ref{eq:9}) finally yields the LEF as a function of the beating signals:

\begin{align}
\mathrm{LEF}=  \bigg|  \frac{B_{1}-B_{-1}}{B_{1}+B_{-1}}  \bigg| ,
\label{eq:10}
\end{align}

\section{Analytical expression for the LEF based on the complex susceptibility}

The spectral response of the laser active medium is fully described with its complex optical susceptibility $\chi = \chi_R + i\chi_I$. Assuming that the gain varies linearly with the carrier density, we can define $\chi(\omega)$ as~\cite{agrawal1995semiconductor}:

\begin{align}
\chi (\omega)=  \frac{\mu^2 \Delta N }{\varepsilon_0 \hbar }  \frac{(1+i\alpha)^2}{\omega - \omega_0 - \frac{i}{T_2}} ,
\label{eq:a1}
\end{align}

 where $\omega$ is the angular frequency, $\omega_0 = 2\pi f_0$, $\mu$ is the dipole matrix element, $\Delta N$ is the carrier density, and $T_2$ is the dephasing time. The value of the LEF at the gain peak is $\alpha$. A non-value $\alpha$-factor describes the asymmetric gainshape of the semiconductor lasers.

The LEF is defined as the ratio of the carrier-induced change of the refractive index and gain. As the optical gain is described by $\chi_I$ and the refractive index is defined with $\chi_R$, we can write $\mathrm{LEF}=-(\nicefrac{\partial \chi_R}{\partial N}) /( \nicefrac{\partial \chi_I}{\partial N})$.
Under the assumption that both the real and imaginary part of the susceptibility are linear functions of the carrier density, we can remove the derivatives in the definition of the LEF~\cite{agrawal1995semiconductor}:

\begin{align}
\mathrm{LEF}(\omega)=-\frac{\chi_R(\omega) }{\chi_I(\omega)}.
\label{eq:a2}
\end{align}

Insertion of Eq.~(\ref{eq:a1}) in Eq.~(\ref{eq:a2}) yields the definition of the spectrally resolved LEF as a function of the value at the gain peak $\alpha$, angular frequency $\omega$ and the dephasing time $T_2$, which describes the gainwidth. It is obtained:

\begin{align}
\mathrm{LEF}(\omega)=\frac{2\alpha - (1-\alpha^2)(\omega-\omega_0)T_2 }{(1-\alpha^2) + 2\alpha(\omega-\omega_0)T_2}.
\label{eq:a3}
\end{align}

It is trivial to confirm that the LEF at the gain peak is equal to the inserted $\alpha$-factor --  $\mathrm{LEF}(\omega_\mathrm{peak}) = \alpha$, where $\omega_\mathrm{peak}=\omega + \alpha/T_2$ is the frequency at the gain peak.

\section{Additional experimental data}

\begin{figure}[th!]
	\centering
	\includegraphics[width = 1 \columnwidth]{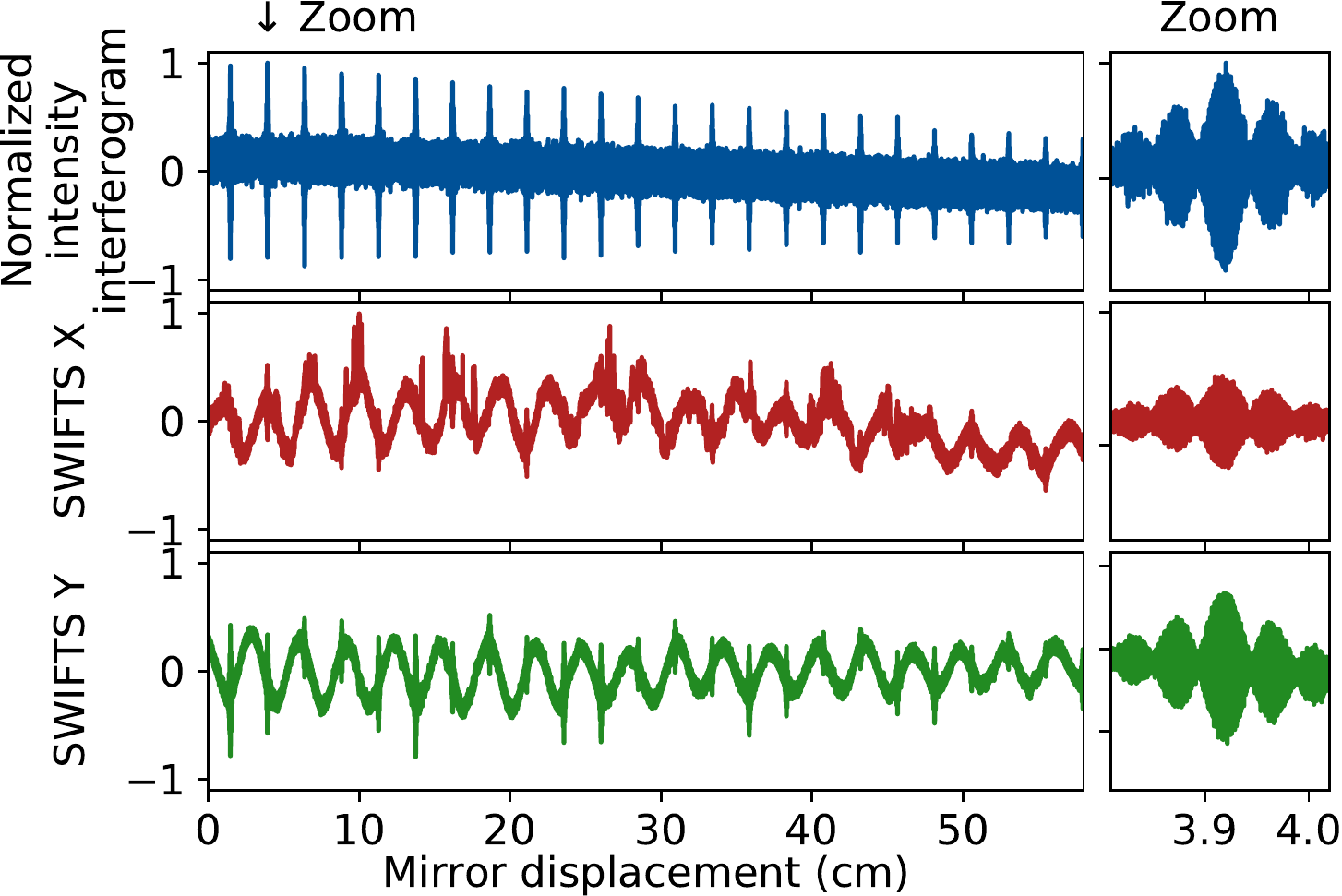}
	\caption{ \textbf{Experimental data of a SWIFTS measurement:} Normalized intensity interferogram (top), Normalized SWIFTS X and Y quadrature interferograms (middle, bottom) with a zoom around the second burst indicated by the arrow (right).}
	\label{suppfig3}
\end{figure}

The recorded intensity interferogram as well as the X and Y trace of the SWIFTS measurement can be seen in Fig.~\ref{suppfig3}. The custom-built FTIR spectrometer allows for a large mirror displacement, resulting in a substantially high resolution in frequency space to clearly distinguish the beatings of the central modes with the weak modulation sidebands.

\end{document}